\documentclass[%
 reprint,
 showpacs,preprintnumbers,
 amsmath,amssymb,
 aps,
 prb,
]{revtex4-2}
\usepackage{dcolumn}
\usepackage{subfigure}
\usepackage{graphicx,color}
\usepackage{mathrsfs}

\makeatletter
\def\btt#1{\texttt{\@backslashchar#1}}
\DeclareRobustCommand\bblash{\btt{\@backslashchar}} \makeatother

\begin{document}

\title{Unconventional enhancement of anomalous Hall effect by tilt of Zeeman field in topologically-nontrivial MXenes, $M_2M'$C$_2$O$_2$
}

\author{Tetsuro Habe}
\affiliation{Department of Mechanical and Electrical Systems Engineering, Kyoto University of Advanced Science, Kyoto 615-8577, Japan}

\date{\today}

\begin{abstract}
In this paper, the anomalous Hall effect of topologically-nontrivial MXenes, $M_2M'$C$_2$O$_2$, and the electronic structure in the presence of magnetic proximity effect is theoretically investigated.
The theoretical analysis is performed in two different ways: an effective model and a multi-orbital tight-binding model generated from the first-principles band structure.
These two theoretical methods provided the similar profile of Berry curvature for electronic states near the bulk band gap, and they show an unconventional rise of hollowed-out peak in the profile with the tilt of proximity magnetic potential.
The anomalous Hall conductivity is also calculated as a function of the charge density and the tilt angle of proximity magnetic order.
Then, an unconventional enhancement of anomalous Hall conductivity by the tilt of magnetic order is theoretically predicted as a result of the variation of Berry curvature.
\end{abstract}

\maketitle
\section{Introduction}

\begin{figure}[htbp]
\begin{center}
 \includegraphics[width=80mm]{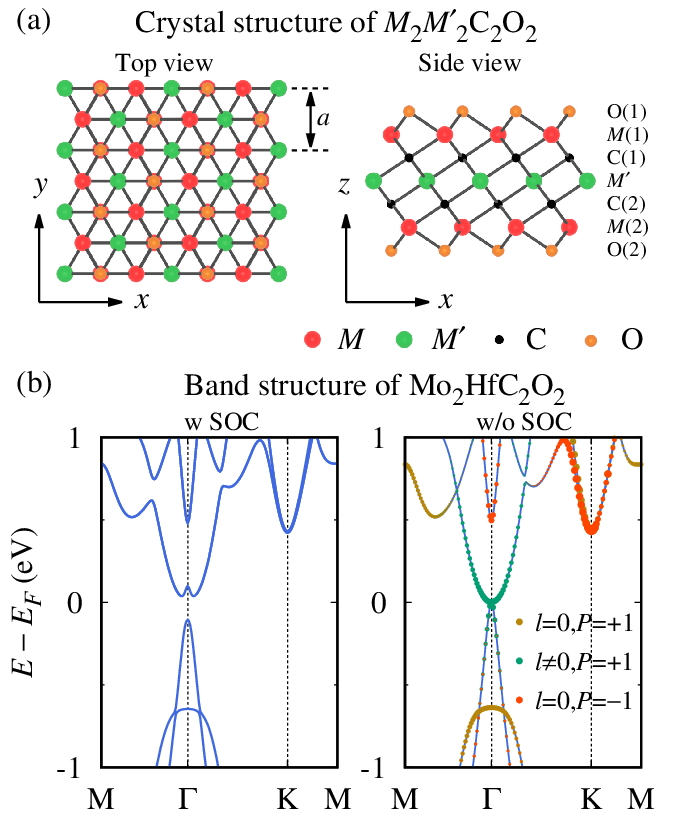}
\caption{The schematic of crystal structure of $M_2M'$C$_2$O$_2$ in (a) and the band structures of Mo$_2$HfC$_2$O$_2$ with and without SOC in (b). The amplitude of Wannier orbitals in Mo with the angular momentum $l_z$ and the parity $P$ are represented by the size of symbols on the band structure without SOC. 
 }\label{fig_band_structure}
\end{center}
\end{figure}

Double-transition-metal carbides are contained in a group of two-dimensional crystals called MXenes.\cite{Anasori2015}
The specific compounds $M_2M'$C$_2$O$_2$ consisting of a group-VI element $M$ (Mo or W) and a group-IV element $M'$ (Ti, Zr, or Hf) are predicted to exhibit topologically-nontrivial electronic properties as time-reversal invariant topological insulators in the presence of oxygen atoms attaching at the surface as shown in Fig.\;\ref{fig_band_structure}(a).\cite{Khazaei2016,Si2016,Huang2020,Parajuli2024}
The topologically-nontrivial electronic structure is generated via the band inversion and the band split due to spin-orbit coupling around the $\Gamma$ point in the Brillouin zone, and they have been theoretically predicted to possess unconventional Berry curvature, a hollowed-out peak structure around the high-symmetry wave number.
\cite{Si2016}
In the presence of time-reversal symmetry, the Berry curvature gives any contribution to electric transport properties except for spin properties because of Kramer's degeneracy.


In the presence of magnetic order, two-dimensional electronic systems exhibit Hall effect, anomalous Hall effect, via the coupling between the magnetic order and the spin degree of freedom.\cite{Haldane2004}
Actually, the anomalous Hall effect has been observed in several experimental set-ups including atomic layered materials\cite{Z.Wang2015,Ng2020,Matsuoka2022} or topological insulators\cite{Alegria2014,Jiang2015,Chi2017,Zhu2018,Mogi2019} on a ferromagnetic substrate which induces a proximity magnetic potential into the materials.
Especially in a material with an intrinsic Berry curvature, the magnetic proximity effect breaks Kramer's degeneracy and exposes the role of the intrinsic Berry curvature.


In this paper, the anomalous Hall effect of the topologically-nontrivial MXenes induced by magnetic proximity effect is theoretically investigated utilizing an analytic model and a multi-orbital tight-binding model.
In Sec.\;\ref{sec_electronic_structure}, the band structure and the relevant atomic orbitals to the electronic states are theoretically given around the Fermi energy. 
In Sec.\;\ref{sec_effective_model}, an effective model is introduced and applied to describe the eigenstates in the materials with and without the Zeeman field representing the magnetic proximity effect.
Then the analytic formula of Berry curvature around the $\Gamma$ point is presented for few bands near the Fermi energy.
In Sec.\;\ref{sec_first-principles}, the electronic states and the Berry curvature are also investigated utilizing a multi-orbital tight-binding model generated from the first-principles band structure for each compound of $M_2M'$C$_2$O$_2$.
In Sec.\;\ref{sec_hall_conductivity}, the Hall conductivity is calculated and presented as a function of the charge density and the tilt angle of Zeeman field.
Finally, the discussion and conclusion are given in Secs.\;\ref{sec_discussion} and \ref{sec_conclusion}, respectively.


\section{Band Structure}\label{sec_electronic_structure}


First, the electronic band structure of double-transition-metal carbides is reviewed around the Fermi energy.
The electronic band calculation is performed in density functional theory (DFT) utilizing QUANTUM ESPRESSO,\cite{Quantum-espresso} a package of numerical codes for DFT calculations with the Predew-Burke-Emzerhof functional.\cite{PBE_functional}
The energy cutoff is 60 Ry for the plane wave basis and 500 Ry for the charge density on the $12\times12\times1$ $k$-mesh in the Brillouin zone, which corresponds to the unit cell given by two primitive lattice vectors, $\boldsymbol{a}_1=(\sqrt{3}a/2,-a/2)$ and $\boldsymbol{a}_2=(0,a)$, with the lattice constant $a$.
The lattice constant $a$, the thickness of layer $c$, and the atomic positions $d_{A-B}$ in the perpendicular direction to the layer are numerically determined within the atomic positions in table \ref{tab_atomic_position} utilizing vc-relax, the lattice optimization code in QUANTUM ESPRESSO, in the conditions of thresholds $10^{-2}$ kbar for stress and $10^{-4}$ Ry/Bohr for forces.
To simulate an isolated two-dimensional monolayer, the distance between adjacent monolayers is fixed at 40{\AA} in DFT calculations.
The convergence criterion is $10^{-8}$ Ry for self-consistent field calculations.


\begin{table}
\caption{The atomic positions in a unit cell for $M_2M'$C$_2$O$_2$ with two transition-metal elements $M$ and $M'$. The coordinate $(x_1,x_2,x_3)$ give the atomic position, $\boldsymbol{r}=x_1\boldsymbol{a}_1+x_2\boldsymbol{a}_2+x_3c\boldsymbol{e}_z$.
}
\begin{ruledtabular}
\begin{tabular}{c c c c c c c c}
&O(1)&$M$(1)&C(1)&$M'$&C(2)&$M$(2)&O(2)\\ \hline
$x_1$&2/3&1/3&2/3&0&-2/3&-1/3&-2/3\\ 
$x_2$&1/3&2/3&1/3&0&-1/3&-2/3&-1/3\\ 
$x_3$&1/2&$d_{M-M'}/c$&$d_{\mathrm{C}-M'}/c$&0&$-d_{\mathrm{C}-M'}/c$&$-d_{M-M'}/c$&-1/2\\ 
\end{tabular}\label{tab_atomic_position}
\end{ruledtabular}
\end{table}


In Fig.\;\ref{fig_band_structure}, the band structure of a double-transition-metal carbide, Mo$_2$HfC$_2$O$_2$, is presented under two conditions: with and without SOC.
For the compound, the optimized crystal parameters are $a=2.990$\AA, $c=7.670$\AA, $d_{M-M'}=2.655$\AA, and $d_{\mathrm{C}-M'}=1.378$\AA.
The two numerical results show that the energy gap is generated by the SOC in the electronic band structure since in the absence of SOC the conduction and valence bands touch to each other at the $\Gamma$ point.
The band structure also indicates that electronic states around the $\Gamma$ point are responsible for the electronic properties for low-level charge doping.
Therefore, an effective model to describe the responsible states is generated in the next section for investigating the Berry curvature.


\section{Effective model analysis}\label{sec_effective_model}


\subsection{Effective model around the $\Gamma$ point}


To generate the effective model, the states at the $\Gamma$ point are characterized in terms of crystal symmetry.
The crystal structure preserves three symmetries: spatial inversion symmetry, threefold rotation symmetry, and mirror symmetry in the $y$ axis in addition to time-reversal symmetry.
Since the $\Gamma$ point is invariant under the symmetric operations, the electronic states have to also be invariant under the operations.
In the absence of SOC, threefold rotation symmetry divides electronic states into three groups possessing different eigenvalues, $R_3=1$, $\omega$, and $\omega^\ast$, for the rotation operator with $\omega=e^{i2\pi/3}$.
Moreover, spatial inversion symmetry classifies each groups into even parity and odd parity, $P=+$ and $P=-$, respectively.
Although the invariance under the mirror operation is also preserved at the $\Gamma$ point, the non-commutativity of threefold rotation and the mirror inversion operations denies the presence of simultaneous eigenstates for these two operations.
In practice, the mirror operation transforms the eigenstate of $\omega$ ($\omega^\ast$) into that of $\omega^\ast$ ($\omega$) unless the state possesses $R_3=1$.
Therefore, the non-commutativity requires degeneracy of electronic states with $\omega$ and $\omega^\ast$ at the $\Gamma$ point.


In the right panel of Fig.\;\ref{fig_band_structure}, the amplitude of $d$ orbitals in Mo atoms are presented on the electronic bands for the characters of the angular momentum and the parity which represent the symmetrical properties of electronic states at the $\Gamma$ point.
The amplitude is calculated utilizing a multi-orbital tight-binding model well-reproducing the first-principles band structure.
The tight-binding model is defined on the basis of $d$-orbitals in two transition-metal elements and $p$-orbitals in C and O.
The hopping parameters are obtained utilizing Wannier90,\cite{wannier90} a numerical code to give the maximally localized Wannier functions and the hopping integrals among them referring to a first-principles band structure.
Since a Wannier orbital on a single atom does not preserve the parity except for the center sublayer of $M'$, the Wannier orbital preserving the parity is implemented superposing the orbitals in the opposite sublayers, e.g., $M(1)$ and $M(2)$ in table \ref{tab_atomic_position}.
The orbital composition reveals that the symmetric character $(R_3,P)$ for the degenerate states at the Fermi energy are $(\omega,+)$ and $(\omega^\ast,+)$, and those of the second top occupied state and the second bottom unoccupied state are $(1,+)$ and $(1,-)$, respectively, at the $\Gamma$ point.
Therefore, electronic states in these band can be represented on the following basis,
\begin{align}
\Psi_\Gamma=(|1,-\rangle,\;|\omega,+\rangle,\;|\omega^\ast,+\rangle,\;|1,+\rangle),\label{eq_basis}
\end{align}
where $|R_3,P\rangle$ indicates a state with $(R_3,P)$.
At the $\Gamma$ point, the Hamiltonian can be given by $\mathrm{diag}[E_-,0,0,E_+]$ on this basis where $E_->0$ and $E_+<0$ are the energies in the second lowest conduction band and the second highest valence band, respectively.


Around the high-symmetry point, the high-symmetric wave functions in Eq.\;(\ref{eq_basis}) are mixed in eigenstates, but the Hamiltonian possesses the invariance among the equivalent wave numbers under each symmetric operation, e.g., $\boldsymbol{k}\rightarrow-\boldsymbol{k}$ for spatial inversion and $(k_x\cos(2\pi/3)-k_y\sin(2\pi/3),\;k_x\sin(2\pi/3)+k_y\cos(2\pi/3))$ for threefold rotation.
The symmetric property derives the effective Hamiltonian up to $k^2$ without SOC,
\begin{align}
\hat{H}_0=
\begin{pmatrix}
\varepsilon_-(\boldsymbol{k})&\alpha{ke^{i\varphi_{\boldsymbol{k}}}}&\alpha{ke^{-i\varphi_{\boldsymbol{k}}}}&0\\
\alpha{ke^{-i\varphi_{\boldsymbol{k}}}}&\varepsilon_0(\boldsymbol{k})&-\gamma k^2e^{-2i\varphi_{\boldsymbol{k}}}&0\\
\alpha{ke^{i\varphi_{\boldsymbol{k}}}}&-\gamma k^2e^{2i\varphi_{\boldsymbol{k}}}&\varepsilon_0(\boldsymbol{k})&0\\
0&0&0&\varepsilon_+(\boldsymbol{k})
\end{pmatrix},\label{eq_original_Hamiltonain}
\end{align}
on the basis in Eq.\;(\ref{eq_basis}) with $\tan\varphi_{\boldsymbol{k}}=k_x/k_y$ and $k=|\boldsymbol{k}|$.
Since the lowest band consists only a single basis $|1,+\rangle$ around the $\Gamma$ point, the band gives no contribution to the anomalous Hall effect in this region.
Thus, the $3\times3$ Hamiltonian without the lowest band,
\begin{align}
\hat{H}=
\begin{pmatrix}
E_-&\alpha(k_x+ik_y)&\alpha(k_x-ik_y)\\
\alpha(k_x-ik_y)&0&-\gamma(k_x-ik_y)^2\\
\alpha(k_x+ik_y)&-\gamma(k_x+ik_y)^2&0
\end{pmatrix},\label{eq_three_orbital_model}
\end{align}
can be utilized for the theoretical analysis of Berry curvature.
Here, the $\boldsymbol{k}$-dependence of the diagonal components is ignored because this simplification does not change the nature of unconventional geometrical phase in the electronic states.
The effective model gives two conduction bands,
\begin{align}
\begin{split}
{E}_2^{(0)}=&E_-+2\Delta_k,\\
{E}_1^{(0)}=&\gamma k^2,\\
\end{split}
\end{align}
and one valence band,
\begin{align}
\begin{split}
{E}_{-1}^{(0)}=&-\gamma k^2-2\Delta_k,
\end{split}
\end{align}
with $\Delta_k$ representing the dispersion of the highest conduction band,
\begin{align}
2\Delta_k=\frac{1}{2}\sqrt{(E_-+\gamma k^2)^2+8\alpha^2k^2}-\frac{1}{2}(E_-+\gamma k^2),
\end{align}
The eigenvectors for these bands are given by
\begin{align}
\begin{split}
|\psi_{2,\boldsymbol{k}}^{(0)}\rangle=&\frac{1}{2}\begin{pmatrix}
\sqrt{2(1+\cos\phi_{k})}\\
\sqrt{1-\cos\phi_{k}}e^{-i\varphi_{\boldsymbol{k}}}\\
\sqrt{1-\cos\phi_{k}}e^{i\varphi_{\boldsymbol{k}}}\\
\end{pmatrix},\\
|\psi_{1,\boldsymbol{k}}^{(0)}\rangle=&\frac{1}{\sqrt{2}}\begin{pmatrix}
0\\
e^{-i\varphi_{\boldsymbol{k}}}\\
-e^{i\varphi_{\boldsymbol{k}}}\\
\end{pmatrix},\\
|\psi_{-1,\boldsymbol{k}}^{(0)}\rangle=&\frac{1}{2}\begin{pmatrix}
-\sqrt{2(1-\cos\phi_{k})}\\
\sqrt{1+\cos\phi_{k}}e^{-i\varphi_{\boldsymbol{k}}}\\
\sqrt{1+\cos\phi_{k}}e^{i\varphi_{\boldsymbol{k}}}\\
\end{pmatrix},
\end{split}\label{eq_basis_wo_soc}
\end{align}
with two phase factors $\varphi_{\boldsymbol{k}}$ and $\phi_{k}$ defined as
\begin{align}
\cos\phi_{k}=&(E_-+\gamma k^2)/\sqrt{(E_-+\gamma k^2)^2+8\alpha^2k^2}.
\end{align}


In the presence of SOC, spatial inversion symmetry prohibits the coupling between orbitals with different parities, and threefold symmetry constrains the direction of the angular momenta coupled, i.e., $\hat{H}_{\mathrm{so}}=\mathrm{diag}[0,-\lambda \sigma_z,\lambda \sigma_z,0]$ on the same basis as Eq.\;(\ref{eq_original_Hamiltonain}) for the up-spin $\sigma_z=1$ and the down-spin $\sigma_z=-1$.
On the basis of Eq.\;(\ref{eq_basis_wo_soc}), the spin-orbit coupling term can be rewritten as
\begin{align}
\hat{H}_{\mathrm{so}}=&-\frac{\lambda \sigma_z}{\sqrt{2}}\begin{pmatrix}
0&\sqrt{1-\cos\phi_{k}}&0\\
\sqrt{1-\cos\phi_{k}}&0&\sqrt{1+\cos\phi_{k}}\\
0&\sqrt{1+\cos\phi_{k}}&0
\end{pmatrix}.
\end{align}
Since the energy difference between two conduction bands $E_-$ is much larger than the spin-orbit coupling energy $\lambda$, the effect of spin-orbit coupling can be ignored in the highest band, i.e., $\cos\phi_{k}\sim1$, and assumed to be constant for the other two bands around the $\Gamma$ point.
Thus the total Hamiltonian on the basis $(|\psi_{2}^{(0)}\rangle,\;|\psi_{1}^{(0)}\rangle,|\psi_{-1}^{(0)}\rangle)$ is given by 
\begin{align}
\hat{H}_{\mathrm{tot}}=\begin{pmatrix}
E_-+2\Delta_k&0&0\\
0&\gamma k^2&\lambda \sigma_z\\
0&\lambda \sigma_z&-\gamma k^2-2\Delta_k
\end{pmatrix}.\label{eq_total_Hamiltonian}
\end{align}
Then the SOC removes the degeneracy at the $\Gamma$ point in the electronic band structure,
\begin{align}
\begin{split}
E_2=&E_-+2\Delta_k\\
E_1=&\lambda_k'-\Delta_k\\
E_{-1}=&-\lambda_k'-\Delta_k,
\end{split}
\end{align}
where $\lambda_k'$ is given by
\begin{align}
\lambda_k'=\sqrt{\lambda^2+(\gamma k^2+\Delta_k)^2}.
\end{align}
Moreover, the eigenstates are represented by linear-combinations of the basis in Eq.\;(\ref{eq_basis_wo_soc}),
\begin{align}
\begin{split}
|\psi_{2,\boldsymbol{k}}^{s_z}\rangle=&|\psi_{2,\boldsymbol{k}}^{(0)}\rangle\\
|\psi_{1,\boldsymbol{k}}^{s_z}\rangle=&\sqrt{\frac{1+\eta_k}{2}}|\psi_{1,\boldsymbol{k}}^{(0)}\rangle+\sigma_{z}\sqrt{\frac{1-\eta_k}{2}}|\psi_{-1,\boldsymbol{k}}^{(0)}\rangle\\
|\psi_{-1,\boldsymbol{k}}^{s_z}\rangle=&\sigma_z\sqrt{\frac{1-\eta_k}{2}}|\psi_{1,\boldsymbol{k}}^{(0)}\rangle-\sqrt{\frac{1+\eta_k}{2}}|\psi_{-1,\boldsymbol{k}}^{(0)}\rangle,
\end{split}\label{eq_basis_w_soc}
\end{align}
for the up-spin $s_z=\uparrow$ and down-spin $s_z=\downarrow$ with a $k$-dependent factor,
\begin{align}
\eta_k=\frac{\gamma k^2+\Delta_k}{\sqrt{(\gamma k^2+\Delta_k)^2+\lambda^2}}.
\end{align}


The Berry curvature along the $z$-axis can be defined via the derivatives of state vector with respect to the wave number,
\begin{align}
\Omega^z_{n}(\boldsymbol{k})=-\mathrm{Im}\left[(\langle\nabla_{\boldsymbol{k}}\psi_{n,\boldsymbol{k}}|\times|\nabla_{\boldsymbol{k}}\psi_{n,\boldsymbol{k}}\rangle)_z\right],
\end{align}
where $|\psi_{n,\boldsymbol{k}}\rangle$ is the Bloch state at $\boldsymbol{k}$ in the $n$-th band.
In the present analytic model, each electronic state can be represented by a linear combination of orthonormal complex vectors at each $\boldsymbol{k}$,
\begin{align}
\boldsymbol{u}=\frac{1}{\sqrt{2}}\begin{pmatrix}
0\\
e^{-i\varphi_{\boldsymbol{k}}}\\
e^{i\varphi_{\boldsymbol{k}}}\\
\end{pmatrix},\;\;
\boldsymbol{v}=\frac{1}{\sqrt{2}}\begin{pmatrix}
0\\
e^{-i\varphi_{\boldsymbol{k}}}\\
-e^{i\varphi_{\boldsymbol{k}}}\\
\end{pmatrix},\;\;
\boldsymbol{w}=\begin{pmatrix}
1\\
0\\
0\\
\end{pmatrix},
\end{align}
as follows,
\begin{align}
|\psi_{n,\boldsymbol{k}}\rangle=f(k)\boldsymbol{u}+g(k)\boldsymbol{v}+h(k)\boldsymbol{w}.
\end{align}
Then, the general formula of Berry curvature is given by
\begin{align}
\Omega_n^z(\boldsymbol{k})=-\frac{2}{k}\mathrm{Re}[(f^\ast g)'],\label{eq_Berry_curvature_1}
\end{align}
where $f'$ is the derivative with respect to $k$.
For the state vectors in Eq.\;(\ref{eq_basis_w_soc}), the above formula concludes the absence of Berry curvature for the second lowest conduction band $|\psi_{2,\boldsymbol{k}}^{s_z}\rangle$ around the $\Gamma$ point,
\begin{align}
\Omega_{2,\sigma_z}^z(\boldsymbol{k})=0.
\end{align}
On the other hand, the Berry curvature for the other two bands $E_n$ with $n=\pm1$ is non-zero and changes its sign with the spin and band indexes,
\begin{align}
\Omega_{n,\sigma_z}^z=s\Omega_0^z(k)\label{eq_simple_form_Berry}
\end{align}
where $s=n\cdot\sigma_z$ is the product of the band index $n=\pm1$ and the spin index $\sigma_z=\pm$.
The wave number dependence is given by
\begin{align}
\Omega_{0}^z(k)
=&-\frac{1}{\sqrt{2}k}\frac{\partial}{\partial k}\left(\frac{\lambda}{\lambda_k'}\sqrt{1+\cos\phi_k}\right).\label{eq_magnitude_Berry}
\end{align}
Around the $\Gamma$ point, the odd-parity band energy $E_2$ is much larger than kinetic energies, i.e.,
\begin{align}
\gamma k^2,\;\alpha k\ll E_-.\label{eq_condition_0}
\end{align}
Then, the Berry curvature is rewritten as
\begin{align}
\Omega^{z}_{0}\simeq\frac{2}{\sqrt{1+(k/\kappa_\lambda)^4}}\left(\frac{\alpha^2}{E_-^2}+\frac{1}{{\kappa_\lambda}^2}\frac{(k/\kappa_\lambda)^2}{1+(k/\kappa_\lambda)^4}\right),\label{eq_exact_form_wo_Zeeman}
\end{align}
with 
\begin{align}
{\kappa_\lambda}^2=\lambda/(\gamma+\alpha^2/E_-).\label{eq_kappa}
\end{align}
The analytic form shows that Berry curvature of each spin band remains to be non-zero even at the $\Gamma$ point.
In the vicinity of the $\Gamma$ point, i.e., $k/\kappa_\lambda\ll1$, its derivative is positive,
\begin{align}
\frac{\partial}{\partial k}\Omega_0^z\simeq\frac{4k}{\kappa_\lambda^4}\left(1-\frac{\alpha^2}{E_-^2}k^2\right),
\end{align}
within the second lowest order of $k$, and thus the Berry curvature increases with $k$ around the $\Gamma$ point.
This unconventional behavior is characteristic in the topologically-nontrivial MXenes.
Although the Berry curvature of each spin band is not zero, the spin-degeneration leads to zero Berry curvature in total for any electronic energy.
The absence of total Berry curvature is a trivial conclusion due to time-reversal invariance, i.e., the absence of magnetic source.


\subsection{Zeeman Field}


In the presence of magnetic proximity effect, the electronic structure loses time-reversal invariance and the degenerated spin states split due to the induced Zeeman field $\boldsymbol{M}/|\boldsymbol{M}|=(\sin \theta\cos q_M,\;\sin \theta\sin q_M,\;\cos\theta)$,
\begin{align}
\hat{U}_Z=-\boldsymbol{M}\cdot\frac{1}{2}\hat{\boldsymbol{\sigma}},\label{eq_original_Zeeman}
\end{align}
with $\hat{\boldsymbol{\sigma}}$ consists of Pauli matrices as $(\hat{\sigma}_x,\hat{\sigma}_y,\hat{\sigma}_z)$ on the spin space.
The Hamiltonian can be represented using $\psi^{s_z}_{j,\boldsymbol{k}}$ by
\begin{align}
\{\hat{U}_Z\}_{ij}=&-
\begin{pmatrix}
M_z(\psi^{\uparrow}_{i,\boldsymbol{k}})^\dagger\psi^{\uparrow}_{j,\boldsymbol{k}}&M_-(\psi^{\uparrow}_{i,\boldsymbol{k}})^\dagger\psi^{\downarrow}_{j,\boldsymbol{k}}\\
M_+(\psi^{\downarrow}_{i,\boldsymbol{k}})^\dagger\psi^{\uparrow}_{j,\boldsymbol{k}}&-M_z(\psi^{\downarrow}_{i,\boldsymbol{k}})^\dagger\psi^{\downarrow}_{j,\boldsymbol{k}}
\end{pmatrix},
\end{align}
with $M_\pm=M_x\pm iM_y$.
Since the state vector in the higher conduction band $E_2$ is orthogonal to those in the other bands, i.e., $(\psi_{2,\boldsymbol{k}}^\sigma)^\dagger\psi_{j,\boldsymbol{k}}^{\sigma'}=\delta_{2,j}$, despite the spin, the Zeeman term remains to be the original form in Eq.\;(\ref{eq_original_Zeeman}) for this band and no inter-band coupling appears.
For the other bands $E_{\pm1}$, on the other hand, the Hamiltonian contains not only the intra-band coupling term,
\begin{align}
\{\hat{U}_Z\}_{jj}=-\begin{pmatrix}
M_z&\eta_kM_-\\
\eta_kM_+&-M_z
\end{pmatrix},
\end{align}
but also the inter-band coupling term,
\begin{align}
\{\hat{U}_Z\}_{1,-1}=-\begin{pmatrix}
0&\sqrt{1-\eta_k^2}M_-\\
-\sqrt{1-\eta_k^2}M_+&0
\end{pmatrix}.
\end{align}
Then the effective Hamiltonian with the Zeeman effect possesses not only the intra-band coupling $\hat{h}_M$ but also the inter-band coupling $\hat{h}_M$,
\begin{align}
\hat{H}_M= \hat{h}_M+\sqrt{1-\eta_k^2}\hat{h}_M'\label{eq_Zeeman_Hamiltonian}
\end{align}
where each component is given by
\begin{align}
\hat{h}_M=&\begin{pmatrix}
E_1-M_z&-\eta_kM_-&0&0\\
-\eta_kM_+&E_1+M_z&0&0\\
0&0&E_{-1}+M_z&-\eta_kM_-\\
0&0&-\eta_kM_+&E_{-1}-M_z\\
\end{pmatrix},\\
\hat{h}_M'=&\begin{pmatrix}
0&0&0&-M_-\\
0&0&M_+&0\\
0&M_-&0&0\\
-M_+&0&0&0
\end{pmatrix},
\end{align}
on the basis of $(|\psi_{1,\boldsymbol{k}}^\uparrow\rangle,|\psi_{1,\boldsymbol{k}}^\downarrow\rangle,|\psi_{-1,\boldsymbol{k}}^\downarrow\rangle,|\psi_{-1,\boldsymbol{k}}^\uparrow\rangle)$.
In the presence of Zeeman field, the spin-degenerate bands split into two branches as follows
\begin{align}
\begin{split}
E^M_{1,\pm}=&-\Delta_k+\sqrt{\left(\lambda_k'\right)^2+M^2\pm 2\lambda'_kM'(\eta_k)},\\
E^M_{-1,\pm}=&-\Delta_k-\sqrt{\left(\lambda_k'\right)^2+M^2\pm 2\lambda'_kM'(\eta_k)},\\
\end{split}\label{eq_Zeeman_energy}
\end{align}
with $M'(\eta_k)=\sqrt{M_z^2+\eta_k^2(M_x^2+M_y^2)}$.
Their electronic states $|\psi_{\pm1,\pm}^M\rangle$ can be the superposition of the spin states in the two bands,
\begin{align}
|\psi^M_{n,\rho}\rangle=c^1_{n,\rho}|\psi_{1}^\uparrow\rangle+c^2_{n,\rho}|\psi_{1}^\downarrow\rangle+c^3_{n,\rho}|\psi_{-1}^\downarrow\rangle+c^4_{n,\rho}|\psi_{-1}^\uparrow\rangle,
\end{align}
and the coefficients $\boldsymbol{c}_{n,\rho}=(c^1_{n,\rho},c^2_{n,\rho},c^3_{n,\rho},c^4_{n,\rho})$ can be given by
\begin{align}
\begin{split}
\boldsymbol{c}_{1,+}=&\begin{pmatrix}
-\sin(\xi_k/2)\cos(\zeta_k^+/2)\\
e^{i\varphi_M}\cos(\xi_k/2)\cos(\zeta_k^+/2)\\
\cos(\xi_k/2)\sin(\zeta_k^+/2)\\
e^{i\varphi_M}\cos(\xi_k/2)\sin(\zeta_k^+/2)\\
\end{pmatrix},\\
\boldsymbol{c}_{1,-}=&\begin{pmatrix}
-\cos(\xi_k/2)\sin(\zeta_k^-/2)\\
-e^{i\varphi_M}\sin(\xi_k/2)\sin(\zeta_k^-/2)\\
-\sin(\xi_k/2)\cos(\zeta_k^-/2)\\
e^{i\varphi_M}\cos(\xi_k/2)\cos(\zeta_k^-/2)\\
\end{pmatrix},\\
\boldsymbol{c}_{-1,+}=&\begin{pmatrix}
\sin(\xi_k/2)\sin(\zeta_k^+/2)\\
-e^{i\varphi_M}\cos(\xi_k/2)\sin(\zeta_k^+/2)\\
\cos(\xi_k/2)\cos(\zeta_k^+/2)\\
e^{i\varphi_M}\sin(\xi_k/2)\sin(\zeta_k^+/2)\\
\end{pmatrix},\\
\boldsymbol{c}_{-1,-}=&\begin{pmatrix}
\cos(\xi_k/2)\cos(\zeta_k^-/2)\\
e^{i\varphi_M}\sin(\xi_k/2)\cos(\zeta_k^-/2)\\
-\sin(\xi_k/2)\sin(\zeta_k^-/2)\\
e^{i\varphi_M}\cos(\xi_k/2)\sin(\zeta_k^-/2)\\
\end{pmatrix},\\
\end{split}
\end{align}
with the two angle parameters $\xi$ and $\zeta^\pm$,
\begin{align}
\tan\xi_k=&\frac{\eta_k\sqrt{M_x^2+M_y^2}}{M_z}=\eta_k\tan\theta_M,\label{eq_angles_Zeeman}\\
\tan\zeta_k^\pm=&\frac{\lambda\sqrt{M_x^2+M_y^2}}{\lambda_k'(\lambda_k'\pm M'(\eta_k))}.
\end{align}
In the presence of the Zeeman field, each bands split into two spin branches in general. However, if the field parallel to the layer, i.e., $M_z=0$, the spin split closes only at the $\Gamma$ point because of $M'(\eta_k)|_{k=0}=0$ in Eq.\;(\ref{eq_Zeeman_energy}).


The Berry curvature for $|\psi_{n,\rho}^M\rangle$ in these spin-split bands changes its sign with the product of the two indexes $s'=n\cdot\rho$ in the similar way of Eq.\;(\ref{eq_simple_form_Berry}),
\begin{align}
\Omega^{z}_{n,\rho}=s'\Omega_{M}^z,
\end{align}
where $\rho=\pm$ is not the spin index in this case, and it indicates the upper branch $\rho=+$ and the lower branch $\rho=-$ of the split band.
Here, $\Omega_{M}^z$ can be also calculated applying the formula in Eq.\;(\ref{eq_Berry_curvature_1}) as follows,
\begin{align}
\Omega_{M}^{z}=-\frac{1}{\sqrt{2}k}\frac{\partial}{\partial k}\left(\frac{\lambda}{\lambda_k'}\sqrt{1+\cos\phi_k}\cos\xi_k\right).\label{eq_exact_form_w_Zeeman}
\end{align}
In the comparison with Eq.\;(\ref{eq_magnitude_Berry}), the effect of Zeeman field appears in the additional factor $\cos\xi_k$ which depends on the angle of the Zeeman field $\boldsymbol{M}$ to the layer as shown in Eq.\;(\ref{eq_angles_Zeeman}).
When the Zeeman field is parallel to the layer, the Berry curvature is absent due to $\cos\xi_k=0$ for any $k$.
For $M_x=M_y=0$, on the other hand, the Berry curvature is equivalent to the form without the Zeeman field in Eq.\;(\ref{eq_magnitude_Berry}).


Under the condition of Eq.\;(\ref{eq_condition_0}) for the effective model, the analytic representation for the Berry curvature can be rewritten as a function of two variables, the angle of the Zeeman field $\theta$ and the normalized wave number $\bar{k}=k/\kappa_\lambda$,
\begin{align}
\begin{split}
\Omega_{M}^z\simeq&\frac{2}{\kappa_\lambda^2}\frac{\cos\theta}{\sqrt{\cos^2\theta+\bar{k}^4}}\left(a_\lambda+\frac{\bar{k}^2}{\cos^2\theta+\bar{k}^4}\right),
\end{split}\label{eq_Berry_curvature_w_Zeeman}
\end{align}
with $a_\lambda=\alpha^2\kappa_\lambda^2/E_-^2$.
Then, the Berry curvature contains only a single parameter $a_\lambda$ in the unit of $2/\kappa_\lambda^2$ and it obviously reproduces $|\Omega_{M}^z=0$ for the field parallel to the layer $\theta=\pi/2$.
Using the same parameters, the energy dispersion of conduction band $E_{1,\rho}^M$ and valence bands $E^M_{-1,\rho}$ can be represented by
\begin{align}
E_{1,\pm}^M=&-\frac{\alpha^2\kappa_\lambda^2}{E_-}\bar{k}^2+\lambda\sqrt{1+\frac{M^2}{\lambda^2}+\bar{k}^4\pm2\frac{M}{\lambda}\sqrt{\cos^2\theta+\bar{k}^4}},\\
E_{-1,\pm}^M=&-\frac{\alpha^2\kappa_\lambda^2}{E_-}\bar{k}^2-\lambda\sqrt{1+\frac{M^2}{\lambda^2}+\bar{k}^4\pm2\frac{M}{\lambda}\sqrt{\cos^2\theta+\bar{k}^4}}.
\end{align}

\begin{figure}[htbp]
\begin{center}
 \includegraphics[width=80mm]{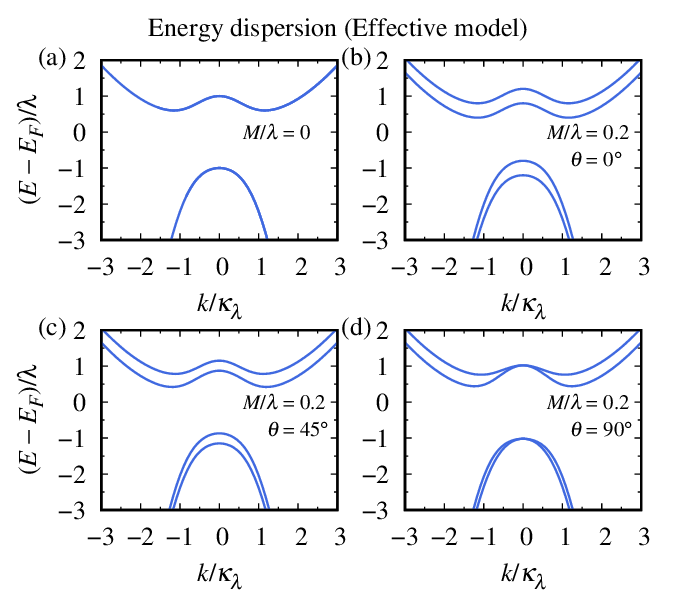}
\caption{The analytically-obtained band structure with/without the Zeeman field around the $\Gamma$ point.
In the horizontal axis, the origin $k=0$ indicates the $\Gamma$ point. The parameters for the calculation are given as follows: $M/\lambda=0.2$ and $(\alpha^2\kappa_\lambda^2/E_-)/\lambda=0.8$. The field direction $\theta$ is defined as the tilt from the perpendicular axis to the layer.
 }\label{fig_Analytic_Bands}
\end{center}
\end{figure}

\begin{figure}[htbp]
\begin{center}
 \includegraphics[width=80mm]{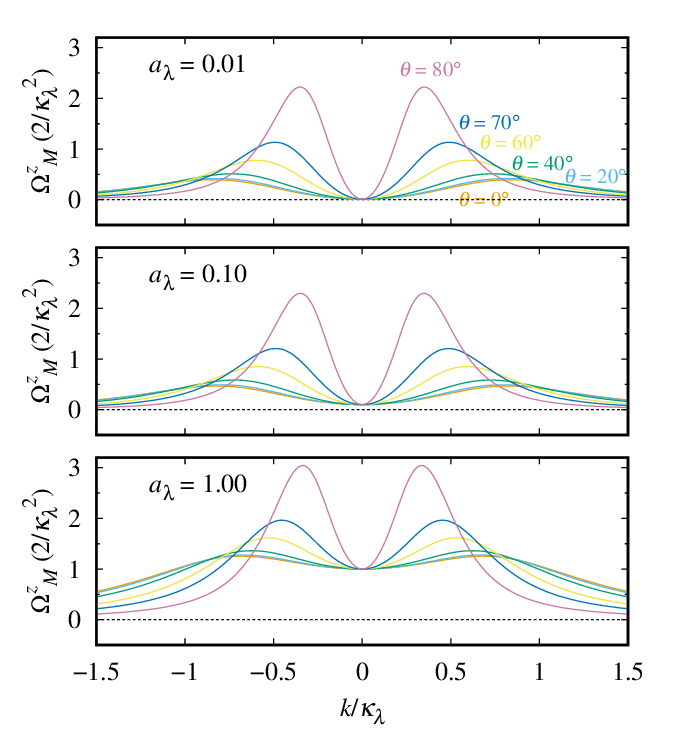}
\caption{The analytically-obtained Berry curvature $\Omega_M^z$ around the $\Gamma$ point.
In each panel, the curves represent the different filed angles $\theta=0^\circ$, $20^\circ$, $40^\circ$, $60^\circ$, $70^\circ$, and $80^\circ$. 
In the horizontal axis, the wave number with respect to the $\Gamma$ point is given in the unit of $\kappa_\lambda$.
 }\label{fig_Analytic_Berry_curvature}
\end{center}
\end{figure}

In Fig. \ref{fig_Analytic_Bands}, the effective band structure is presented for several tilted Zeeman fields to the layer with the angle $\theta=0^\circ$, $45^\circ$, and $90^\circ$.
The pristine band structure in (a) well reproduces the dispersion of the first-principles band in Fig.\;\ref{fig_band_structure} around the Fermi energy.
In the presence of Zeeman field, the spin degenerate band splits into two branches as shown in (b), (c), and (d).
Especially for $\theta=90^\circ$, the two branches are attached only at the $\Gamma$ point.
The effective Berry curvature is also presented in Fig. \ref{fig_Analytic_Berry_curvature} as a function of the wave number $k$ and the angle of Zeeman field $\theta$.
The results show that the Berry curvature increases with the wave number $k$ and gives the maximum at $k\neq0$ for any $\theta$ except  for $90^\circ$. 
Moreover, the amplitude of the peak is enhanced by the tilt of field and the position closes to the $\Gamma$ point, i.e., $k=0$.
The amplitude at $k=0$ varies with the material parameters via $a_\lambda$.
Since the amplitude of Berry curvature is given in the unit of $2/\kappa_\lambda^2\propto\lambda^{-1}$, it is suppressed in a compound with a strong SOC, i.e., the energy gap, roughly.
In the next section, the validity of the analytic results is confirmed comparing it with the more realistic numerical results given by the multi-orbital tight-binding model.


\section{Multi-orbital Tight-binding Model Analysis}\label{sec_first-principles}


\begin{figure}[htbp]
\begin{center}
 \includegraphics[width=80mm]{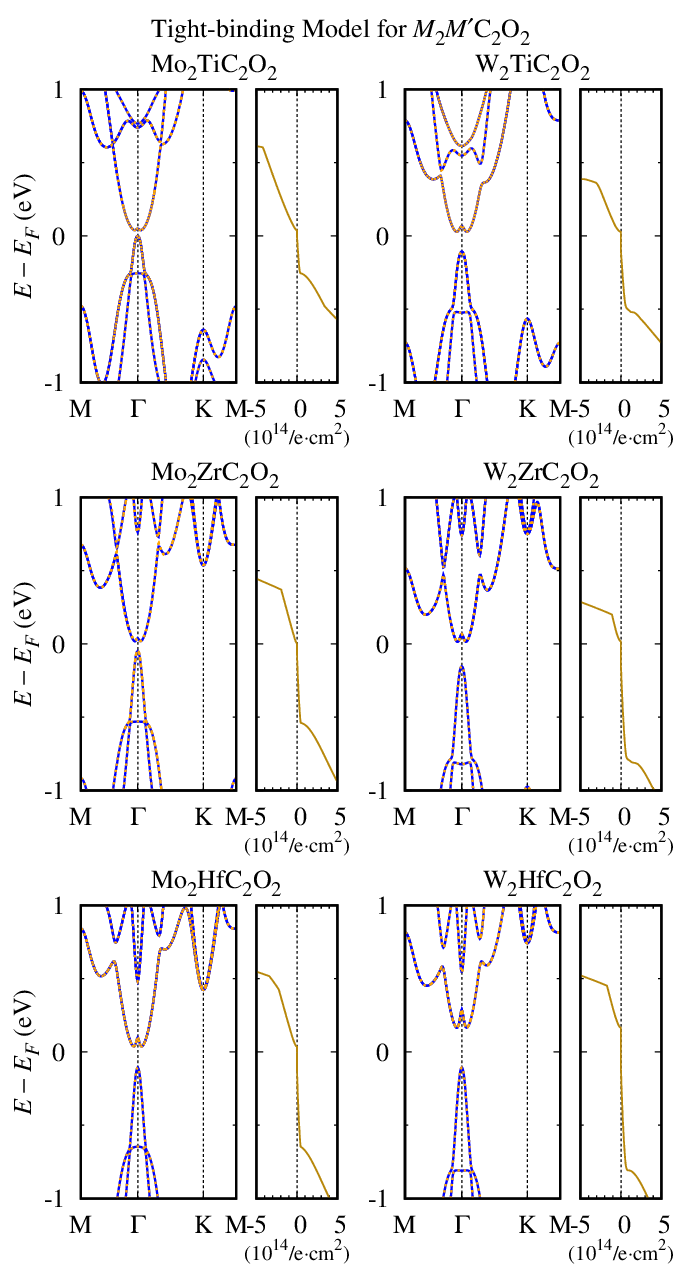}
\caption{The band structure in the left panel and the charge density in the right panel for $M_2M'$C$_2$O$_2$ with SOC. In the left panel, the dashed lines present the bands obtained using the tight-binding model on the first-principles bands.
 }\label{fig_tight-binding_bands}
\end{center}
\end{figure}

To investigate actual MXenes, the effect of Zeeman field and the Berry curvature are theoretically analyzed utilizing a multi-orbital tight-binding model for each $M_2M'$C$_2$O$_2$.
The theoretical model is defined on the basis consisting of $d$-orbitals in $M$ and $M'$, and $p$-orbitals in C and O, and its hopping integrals among the orbitals are numerically obtained utilizing Wannier90\cite{wannier90} as explained in Sec.\;\ref{sec_effective_model}.
The SOC is implemented in the hopping integrals due to the use of the fully-relativistic first-principles band structures as the references for the calculations.
In Fig.\;\ref{fig_tight-binding_bands}, the band structure obtained in the different calculation methods: the first-principles calculation and tight-binding model, is presented for comparison.
The coincidence of the band structures in the different ways shows the validity of the tight-binding models to the theoretical simulations of electronic properties of actual $M_2M'$C$_2$O$_2$.
In the right panels, the charge density $n_c$ is presented for the sift of Fermi energy.
It is numerically evaluated by the following formula,
\begin{align}
n_c(E)=n_n-e\sum_{n}\int_{\mathrm{BZ}}\frac{d^2\boldsymbol{k}}{(2\pi)^2}\theta (E-E_{n\boldsymbol{k}}),
\end{align}
where $\theta(E)$ and $n_n$ are the step function and the nucleus charge density, respectively, with the elementary charge $e$.
The numerical results show that the feasible doping level $\sim10^{14}/e\cdot\mathrm{cm}^2$ in atomic layered materials (see Ref. \onlinecite{Y.Zhang2012} for example) can be achieved for the Fermi energy from -1\;eV to 0.5\;eV.
Therefore, the two valence bands and one conduction band for each spin are responsible for electronic transport phenomena in realistic experimental situations.


\begin{figure}[htbp]
\begin{center}
 \includegraphics[width=80mm]{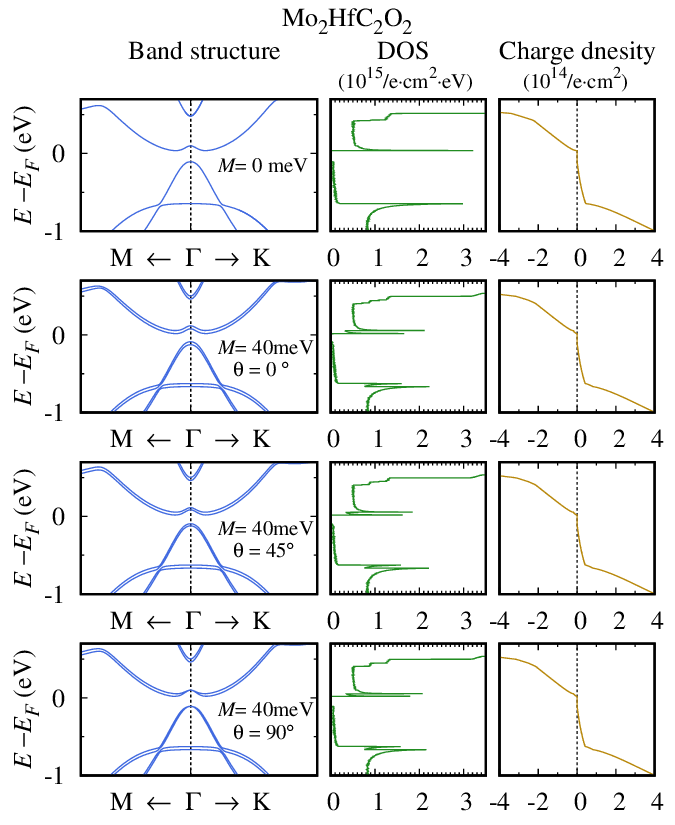}
\caption{The electronic structure of Mo$_2$HfC$_2$O$_2$ in the absence and presence of the Zeeman field with the amplitude $M=40$meV. The leftmost panels present the effect of the Zeeman field on the band structure with the field direction $\theta$ to the $z$ direction. The density of states (DOS) and the charge density are presented in the middle and rightmost panels, respectively.
 }\label{fig_bands_w_Zeeman_effect}
\end{center}
\end{figure}

The Zeeman effect on the materials is investigated utilizing the tight-binding model with an on-site potential of the product of the spin operators $\hat{\boldsymbol{s}}=(\hat{\sigma}_x/2,\hat{\sigma}_y/2,\hat{\sigma}_z/2)$ and the constant vector $\boldsymbol{M}$ for each orbital.
Here, the vector $\boldsymbol{M}$ represents the amplitude and direction of Zeeman field same as that in the effective model of Eq.\;(\ref{eq_original_Zeeman}).
In the presence of the field $\boldsymbol{M}$, the spin-degenerate bands are split into two branches as shown in Fig.\;\ref{fig_bands_w_Zeeman_effect}.
Here, the direction of the field is indicated by the angle $\theta$ to the $z$-axis in the $yz$-plane.
Moreover, the density of states (DOS) is defined as the derivative of the electron density,
\begin{align}
D(E)=\frac{d}{d E}\left(-\frac{n_c}{e}\right).
\end{align}
The spin-split can be observed in the wave number space except for the $\Gamma$ point with $\theta=90^\circ$, and it also appears in the DOS spectrum as a split of peak structure.
The variation of the band structure near the Fermi energy is qualitatively equivalent to that given by the effective model analysis in Fig.\;\ref{fig_Analytic_Bands}.


\begin{figure}[htbp]
\begin{center}
\includegraphics[width=80mm]{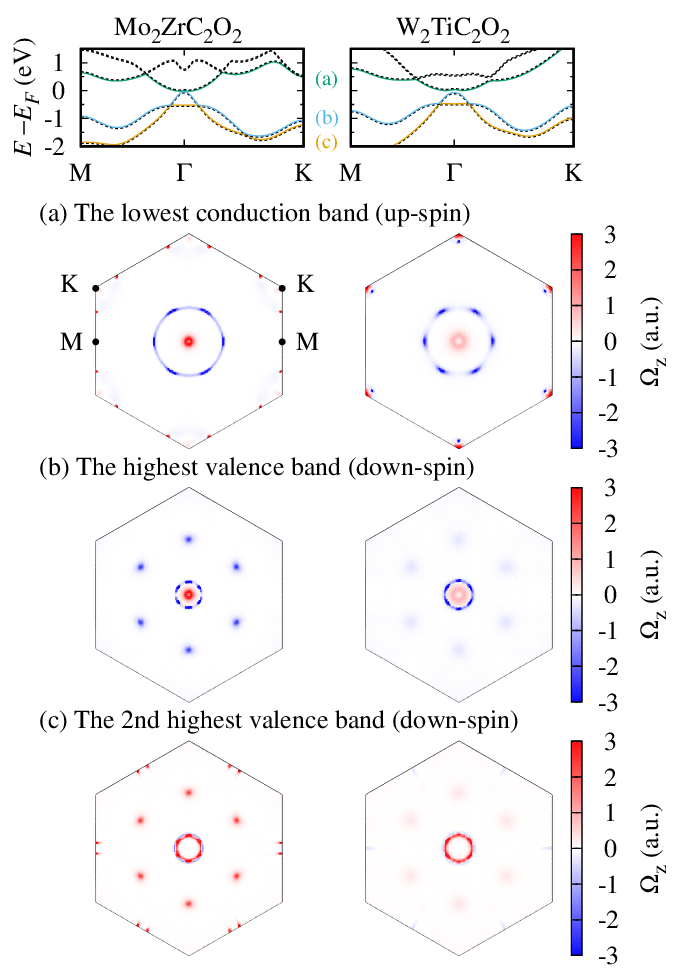}
\caption{The Berry curvature in electronic bands around the Fermi energy for Mo$_2$ZrC$_2$O$_2$ (left panels) and W$_2$TiC$2$O$_2$ (right panels). For each band, one of the spin-split branch is adopted since the two branches exhibit the Berry curvature with the opposite spin. The adopted branches are represented by the solid lines in the top panels. 
 }\label{fig_Berry_curvature}
\end{center}
\end{figure}

For the spin-split bands, the Berry curvature can be calculated utilizing the formula,\cite{Yao2004}
\begin{align}
\Omega^z_n(\boldsymbol{k})=-\sum_{m\neq n}\frac{2\mathrm{Im}[\langle n\boldsymbol{k}|\hat{v}_x|m\boldsymbol{k}\rangle\langle m\boldsymbol{k}|\hat{v}_y|n\boldsymbol{k}\rangle]}{(\omega_{m\boldsymbol{k}}-\omega_{n\boldsymbol{k}})^2},
\end{align}
where $\hat{v}_\mu$ is the velocity operator and $|m\boldsymbol{k}\rangle$ is the state vector at the wave vector $\boldsymbol{k}$ in the $m$-th band $E_{m\boldsymbol{k}}=\hbar\omega_{m\boldsymbol{k}}$.
The Berry curvature is calculated for relevant bands in the hexagonal Brillouin zone and presented for Mo$_2$ZrC$_2$O$_2$ and W$_2$TiC$_2$O$_2$ in Fig.\;\ref{fig_Berry_curvature}. 
In the lowest conduction band, the large negative Berry curvature is distributed along a closed curve enclosing the positive large spot at the $\Gamma$ point.
The positive spot corresponds to the Berry curvature obtained by the effective model in Eq.\;(\ref{eq_Berry_curvature_w_Zeeman}). 
On the other hand, the closed trajectory giving the negative curvature is attributed to the electronic structure apart from the Fermi energy and appears along the wave numbers at which the two conduction bands are close to each other.
The large Berry curvature is explained as the resulting from the band repulsion at the intersection of two bands.\cite{Habe2017,Du2018,Matsuoka2022}
On this curve, the amplitude of Berry curvature is enhanced especially along the $\Gamma$-M direction where the gap between conduction bands is minimized as shown in Fig.\;\ref{fig_tight-binding_bands}.
The trajectory possessing a large Berry curvature is also observed along the trajectory giving a small gap originated to the intersection of two valence bands as shown in (b) and (c) of Fig.\;\ref{fig_Berry_curvature}.
Thus, the equivalent trajectories appear in the two valence bands and gives the Berry curvature with the opposite sign.


\begin{figure}[htbp]
\begin{center}
 \includegraphics[width=80mm]{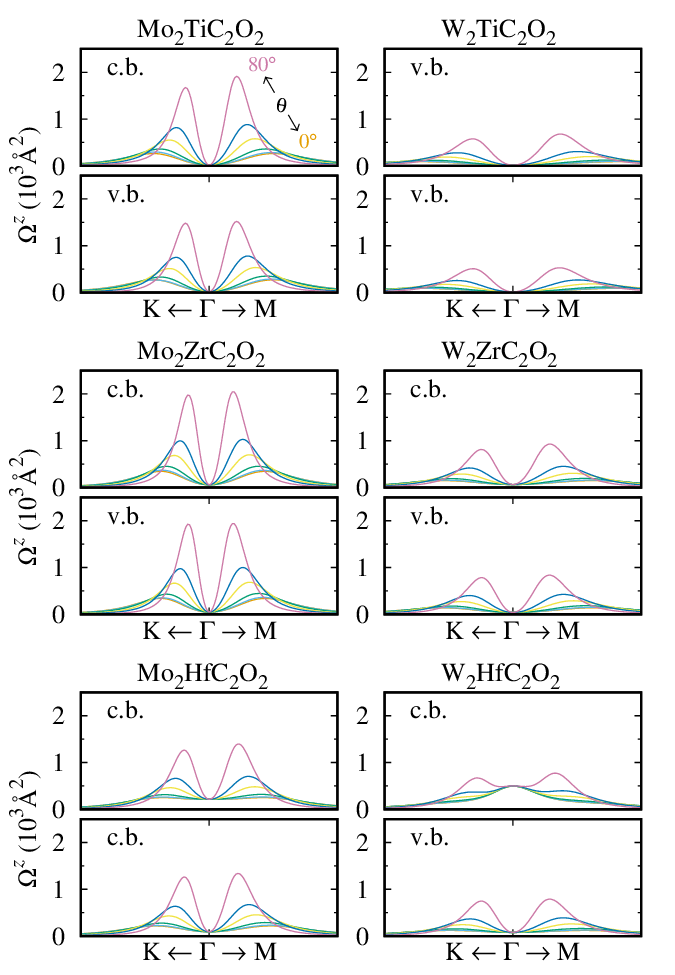}
\caption{The Berry curvature around the $\Gamma$ point for the lowest conduction band (c.b.) and the highest valence band (v.b.).
In each panel, the curves for the different filed angles $\theta=0^\circ$, $20^\circ$, $40^\circ$, $60^\circ$, $70^\circ$, and $80^\circ$ are presented. 
In the horizontal axis, the left and right directions indicates the paths from the $\Gamma$ point to the $K$ and $M$ points, respectively.
 }\label{fig_1D_Berry_curvature}
\end{center}
\end{figure}

To investigate the property for a low charge density, the variations of Berry curvature around the $\Gamma$ point is presented along the $K-\Gamma$ and $\Gamma-M$ lines for the lowest conduction band and the highest valence band of $M_2M'$C$_2$O$_2$ in Fig.\;\ref{fig_1D_Berry_curvature}.
For all compounds, the Berry curvature possesses hollowed-out peaks around the $\Gamma$ point.
The peak positions get close to the $\Gamma$ point and the peak width changes to be narrow with the tilt angle.
Especially in terms of amplitude, the tilt of Zeeman field enhances it around the $\Gamma$ point.
The unconventional enhancement implies the increase of Hall conductivity with the tilt angle in the low charge density region.
These characteristic angle-dependent properties can be also found in the effective theory and reproduced in Fig.\;\ref{fig_Analytic_Berry_curvature}.


Beside the common features over the compounds, the Berry curvature exhibits the dependence on the elements.
The compound with a heavier element possesses the smaller peak amplitude.
Since the heavy element induces the large SOC in general, the suppression of the peak amplitude is consistent with the conclusion of the effective model analysis.
Moreover, the amplitude at the $\Gamma$ point also shows the compound dependence same as that of the result from the effective model in Eq.\;(\ref{eq_Berry_curvature_w_Zeeman}) via the parameter $a_\lambda$ characterizing the material.
Therefore, the numerical results of Berry curvature indicate that the effective model in Eq.\;(\ref{eq_total_Hamiltonian}) well describes the electronic structure of topologically-nontrivial MXenes, $M_2M'$C$_2$O$_2$, including the internal degrees of freedom.


\section{Anomalous Hall conductivity}\label{sec_hall_conductivity}

\begin{figure}[htbp]
\begin{center}
 \includegraphics[width=80mm]{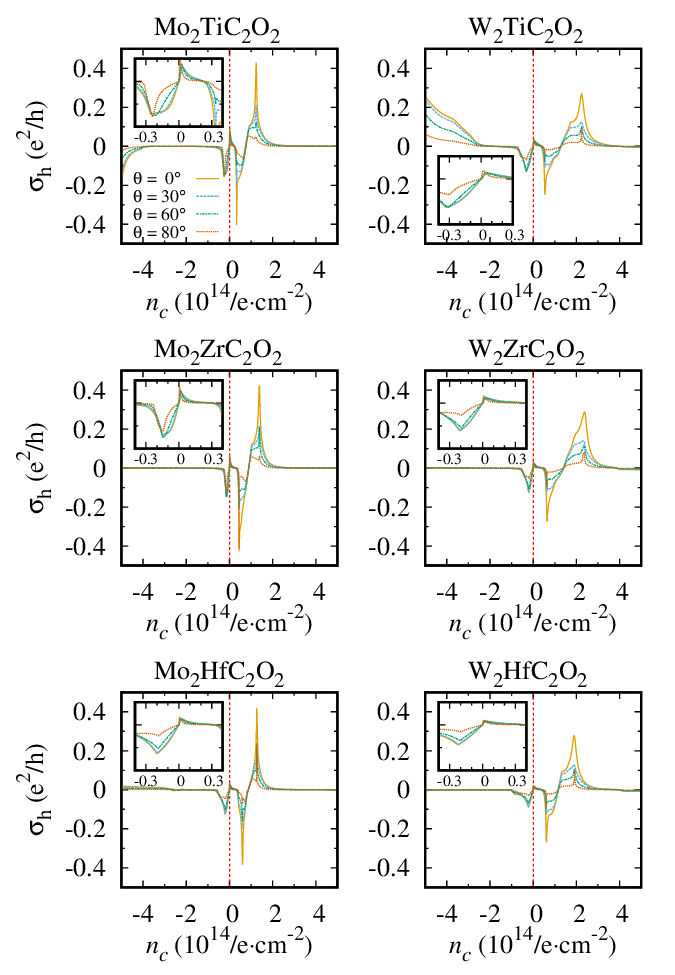}
\caption{The anomalous Hall conductivity for topologically-nontrivial MXenes, $M_2M'$C$_2$O$_2$. The horizontal axis indicates the charge density and the vertical axis is given in the unit of $e^2/h$. The numerical results for different angles of Zeeman field with the amplitude $M=40$meV are presented in each panel. The behavior around the charge neutral point is presented in insets.
 }\label{fig_Hall_conductivity}
\end{center}
\end{figure}

Finally, the anomalous Hall conductivity $\sigma_h$ is calculated utilizing the multi-orbital tight-binding model as the observable associated with the Berry curvature.
The Hall conductivity can be obtained as the summation of the Berry curvature over occupied states,
\begin{align}
\sigma_h(E)=-\frac{e^2}{\hbar}\int\frac{d^2\boldsymbol{k}}{(2\pi)^2}\Omega^z_n(\boldsymbol{k})\theta(E-E_{n\boldsymbol{k}}).\label{eq_Hall_conductivity}
\end{align}
In Fig.\;\ref{fig_Hall_conductivity}, the numerically obtained Hall conductivity is presented as a function of charge density with a specific amplitude of Zeeman field $M=40$ meV for topologically-nontrivial MXenes, $M_2M'$C$_2$O$_2$.
The positive and negative charge densities indicate the shift of the Fermi energy into the valence and conduction bands, respectively.
For the Fermi energy slightly shifting to the conduction band, the lowest branch is occupied partially, and thus the Berry curvature in Fig.\;\ref{fig_Berry_curvature}(a) contributes to the Hall conductivity with the minus sign according to Eq.\;(\ref{eq_Hall_conductivity}).
On the other hand, the doped holes shift the Fermi energy into the valence band, and thus the highest branch possesses unoccupied states above the Fermi energy.
Therefore, the Berry curvature corresponding to the unoccupied states in Fig.\;\ref{fig_Berry_curvature}(b) contributes to the Hall conductivity without sign change.
Actually, the sign of Hall conductivity is positive with a slight hole doping and negative with a slight electron doping.
The amplitude of Hall conductivity is given in the unit of $e^2/h$, the quanta of conductivity.
Since the anomalous Hall effect is attributed to a single band and a small part of another in the range of charge density, the amplitude is less than unity in this unit.

In all compounds, the sign change and opposite peak structures of Hall conductivity can be observed around the charge neutral point and a positive charge density $n_c\sim1\times10^{14}/e\mathrm{\cdot cm}^2$.
These two characteristic charge densities correspond to the energies laying between two bands.
Actually, the hole density $n_c\sim1-2\times10^{14}/e\cdot$cm$^2$ between the large peaks corresponds to the shift of Fermi energy close to the closed trajectory with small gap between two valence bands providing a large amplitude of Berry curvature as shown in Fig.\;\ref{fig_Berry_curvature}(b) and (c).
Especially in W$_2$TiC$_2$O$_2$, a large Hall conductivity can be also obtained with electron-doping which shifts the Fermi energy close to the small gap between the lowest and second lowest conduction bands.
The small gap between two conduction bands can be also observed in the other compounds, but local minima of the conduction band provides possess a large contribution to the DOS and prevent the sift of Fermi energy to the source of large Berry curvature within the feasible charge-doping level.
These peaks of Hall conductivity apart from the charge neutral point exhibit the conventional tilt-angle dependence of the Zeeman field, i.e., the monotonic decrease with $\theta$. 
However, the peaks around the charge neutral point show the unconventional behavior as a function of the tilt angle $\theta$.


\begin{figure}[htbp]
\begin{center}
 \includegraphics[width=80mm]{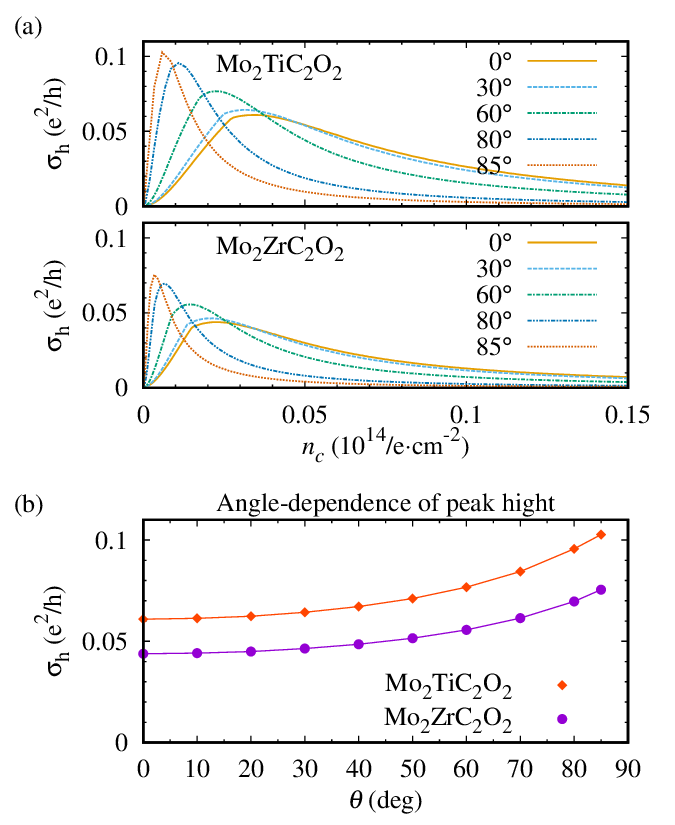}
\caption{The anomalous Hall conductivity with a small hole density. In (a) the hole density dependence is presented for several angles of the Zeeman field in the case of Mo$_2$ZrC$_2$O$_2$. In (b), the angle dependence of peak value is given for Mo$_2$TiC$_2$O$_2$ and Mo$_2$ZrC$_2$O$_2$, which possess the large Hall conductivity.
 }\label{fig_angle_dependence_of_peaks}
\end{center}
\end{figure}

The variation of Hall conductivity near the charge neutral point is unconventional with the tilt of Zeeman field as shown in the insets of Fig.\;\ref{fig_Hall_conductivity}.
Especially in the hole-doped region, the peak height is raised with the tilt angle $\theta$ for all the compounds.
In Fig.\;\ref{fig_angle_dependence_of_peaks}, the angle dependence of peak structure is shown for two compounds to possess large peak heights, Mo$_2$TiC$_2$O$_2$ and Mo$_2$ZrC$_2$O$_2$.
With the tilt of Zeeman field, the peak height and position are raised and shifted to the charge neutral point, respectively, as shown in Fig.\;\ref{fig_angle_dependence_of_peaks}(a).
The increase of the maximum value is also presented in Fig.\;\ref{fig_angle_dependence_of_peaks}(b).
These results indicate that one can enhance the Hall conductivity tilting the direction of Zeeman field to the layer and obtain a large value, e.g., 1.7 times as much as that for $\boldsymbol{M}=(0,0,M_z)$.


Although the maximum value of Hall conductivity is enhanced by the tilt of Zeeman field, the peak width narrows according to $\theta$.
Since the Hall conductivity vanishes for $\theta=90^\circ$, the width decreases to zero with $\theta$.
Then, it is hard to tune the charge density for giving the large peak value with the tilted field almost parallel to the layer.
However, the angle around $\theta=80^\circ-85^\circ$ seams to be feasible in actual experimental situations for other atomic layered materials.\cite{Y.Zhang2012,Chu2014,Joo2016}


\section{Discussion}\label{sec_discussion}


In this section, the unconventional enhancement of anomalous Hall conductivity by the tilt of Zeeman field $\theta$ is discussed in terms of the electronic structure $M_2M'$C$_2$O$_2$.
Since the conventional two-dimensional magnetic materials exhibit the Hall conductivity proportional to the perpendicular magnetic moment, i.e., $\sigma_h\propto M_z=M\cos\theta$,\cite{Nagaosa2010,Sabzalipour2019} the Hall conductivity monotonically decreases with the angle of the Zeeman field $\theta$.
In this work, it is shown that the topologically-nontrivial MXenes contrarily enhances the Hall conductivity with the tilt angle.
As clearly shown in Figs.\;\ref{fig_Analytic_Berry_curvature} and \ref{fig_1D_Berry_curvature}, the enhancement is attributed to the magnetic angle-dependence of Berry curvature peak around the $\Gamma$ point.
The peak structure is well described by the analytic formula in Eq.\;(\ref{eq_Berry_curvature_w_Zeeman}) and the original derivative form in Eq.\;(\ref{eq_exact_form_w_Zeeman}) based on the three-orbital model given in Eq.\;(\ref{eq_three_orbital_model}).
The original form contains the features of the materials via two factors, i.e., the deviation of SOC gap $\lambda/\lambda_k$ and the mixing ratio $\cos\phi_k$ of the odd-parity state $|1,-\rangle$ and the even-parity states $|\omega(\omega^\ast),+\rangle$ as represented in Eq.\;(\ref{eq_basis_wo_soc}).
Thus, the coexistence of these states around the Fermi energy is necessary for the unconventional behavior of Berry curvature.


In conventional monolayer materials, e.g., Graphene\cite{Neto2009} and transition-metal dichalcogenides,\cite{Xiao2012} since the sublayer degree of freedom is absent or irrelevant, the orbitals possess the same parity around the Fermi energy and thus they cannot show the unconventional Berry curvature.
In the topologically-nontrivial MXenes, on the other hand, the transition-meal sublayers $M(1)$ and $M(2)$ possess a major contribution to the electronic structure around the Fermi energy as shown in Fig.\;\ref{fig_band_structure}, and thus they can provide the relevant orbitals with different parities in Eq.\;(\ref{eq_basis}) around the Fermi energy. 
Moreover, threefold rotation and spatial parity symmetries are also responsible for the unconventional enhancement of anomalous Hall conductivity by the tilt of Zeeman field.


\section{Conclusion}\label{sec_conclusion}


In conclusion, the topologically-nontrivial double-transition-metal carbides, $M_2M'$C$_2$O$_2$, are theoretically investigated in terms of nontrivial phase of electronic states, Berry curvature, and predicted to exhibit anomalous Hall effect in the presence of Zeeman field.
The large Hall conductivity can be obtained for the feasible region of doped carriers according to the numerical calculation based on the first-principles band structures.
Moreover, the unconventional enhancement of Hall conductivity by the tilt of Zeeman field is also predicted for a small amount of hole doping.
The unconventional behavior of Hall conductivity is attributed to the rise of hollowed-out peak of Berry curvature with the tilt angle.
The theoretical investigation utilizing three-orbital analytic model well reproduces the variation of Berry curvature and reveals that the origin of the variation is the orbital-set around the Fermi energy due to the sublayer degree of freedom and the fundamental symmetries of the materials.


\begin{acknowledgements}
This work was supported by JSPS KAKENHI Grant Number JP23K03289.
\end{acknowledgements} 

\bibliography{Topological_MXene}

\end{document}